\documentclass[preprint,12pt]{elsarticle}

\usepackage{amssymb}
\usepackage{lineno}
\linenumbers
\usepackage{caption}
\usepackage{subcaption}
\title{Electron-beam Calibration of Aerogel Tiles for the HELIX RICH Detector}
\author[osu]{P.~Allison}
\author[quu]{M.~Baiocchi}
\author[osu]{J.~J.~Beatty}
\author[uch]{L.~Beaufore}
\author[osu]{D.~H.~Calderone}
\author[psu]{Y.~Chen}
\author[psu]{S.~Coutu}
\author[mcg]{E.~Ellingwood\fnref{fn1}}
\author[umi]{N.~Green}
\author[mcg]{D.~Hanna\corref{cor1}} 
\author[uch]{H.~B.~Jeon}
\author[uch]{R.~Mbarek\fnref{fn2}} 
\author[osu]{K.~McBride}
\author[psu]{I.~Mognet}
\author[inu]{J.~Musser}
\author[nku]{S.~Nutter}
\author[mcg]{S.~O'Brien\fnref{fn3}}
\author[quu]{N.~Park}
\author[mcg]{T.~Rosin}
\author[cba]{M.~Tabata}
\author[umi]{G.~Tarl\'e}
\author[inu]{G.~Visser}
\author[uch]{S.~P.~Wakely}
\author[psu]{M.~Yu\fnref{fn5}}
\affiliation[osu]{The Ohio State University, Department of Physics, Columbus, OH, USA}
\affiliation[uch]{University of Chicago, Department of Physics,  Chicago, Il, USA}
\affiliation[psu]{Pennsylvania State University, Department of Physics, University Park, PA,  USA}
\affiliation[inu]{Indiana University, Department of Physics,  Bloomington, USA}
\affiliation[umi]{University of Michigan, Department of Physics, Ann Arbor, MI, USA}
\affiliation[mcg]{McGill University, Department of Physics,  Montreal, QC, Canada}
\affiliation[nku]{Northern Kentucky University, Department of Physics, Geology and Engineering Technology, Highland Heights, KY, USA}
\affiliation[quu]{Queen's University, Department of Physics, Engineering Physics and Astronomy, Kingston, ON Canada}
\affiliation[cba]{Chiba University, Department of Physics, Chiba, Japan}

\cortext[cor1]{Corresponding author}
\fntext[fn1]{Now at Queen's University, Kingston, ON, Canada}
\fntext[fn2]{Now at Joint Space-Science Institute, University of Maryland, College Park, MD, USA}{}
\fntext[fn3]{Also at Arthur B. McDonald Institute, Kingston, ON, Canada}
\fntext[fn5]{Now at Institut de Física d’Altes Energies (IFAE),  Barcelona, Spain}

\begin{document}

\begin{frontmatter}

\begin{abstract}
%% Text of abstract
The HELIX cosmic-ray detector is a balloon-borne instrument designed to measure the flux of light isotopes in the energy range from 0.2 GeV/n to beyond 3 GeV/n. It will rely on a ring-imaging Cherenkov (RICH) detector for particle 
identification at energies greater than 1 GeV/n and will use aerogel tiles with refractive index near 1.15 as the radiator. To achieve the performance goals of the experiment it is necessary to know the refractive index and its position dependence over the lateral extent of the tiles to a precision of O($10^{-4}$). In this paper we describe the apparatus and methods developed to calibrate the HELIX tiles in an electron beam, in order to meet this requirement.
\end{abstract}
\end{frontmatter}

\section{Introduction}

The High Energy Light Isotope eXperiment (HELIX)~\cite{allison} is being developed to measure the chemical and isotopic abundances of light cosmic-ray nuclei. 
The primary goal is to measure the $^{10}$Be/$^9$Be ratio over the energy range from 0.2 GeV/n to beyond 3 GeV/n. 
The detector uses a one-Tesla superconducting magnet with a drift-chamber tracker for measuring particle rigidities. 
A system of time-of-flight (TOF) scintillation detectors is used to measure velocities at energies up to 1 GeV/n and a ring-imaging Cherenkov (RICH) detector is used at higher energies. By combining its charge, measured with the TOF detectors, with its mass, calculated from the rigidity, and velocity, one can uniquely identify each incident particle.

The RICH~\cite{wisher} comprises a 600 mm $\times$ 600 mm radiator plane and a 1000 mm $\times$ 1000 mm detector plane located 500 mm below.
The detector plane is populated with silicon-photomultiplier pixels, each 6 mm $\times$ 6 mm in area and on a grid with 6.2 mm pitch. 
They are deployed, for the first HELIX flight, in a checkerboard pattern that uniformly samples 50\% of the detector plane. 
The radiator plane is made from 36 tiles, each 100 mm $\times$ 100 mm.
Four of the tiles are made of NaF, with a refractive index of 1.33 at 400 nm, and the remaining tiles are made of hydrophobic silica aerogel, approximately 10 mm thick, with a refractive index of approximately 1.15 at 400 nm.

The aerogel tiles were produced at Chiba University~\cite{tabata, tabata-3} using a pinhole drying technique to achieve the relatively high value for the refractive index.
After drying, which involves considerable shrinking, the tiles were approximately 116 mm on a side so they were trimmed to the required 100 mm using a water-jet cutter.
Manufacturing tolerances are such that small variations in aerogel density, as well as tile thickness and surface flatness, are to be expected.
These variations can affect the velocity resolution of the RICH and must be understood and corrected for in the data analysis.
The density variations are the most important, as the density is correlated with the refractive index of the aerogel. 
To obtain optimum results from the HELIX RICH detector we need to produce a map of the refractive index as a function of lateral position in each of the HELIX tiles. 
We do not attempt to measure the index value as a function of depth within the tile since it is not possible using this method.
It is also not straightforward to implement any knowledge of its depth variation in the analysis of flight data. 
This is not expected to be a problem since the results of the study reported here show that the lateral variation of the index is slow and this implies that 
using a constant value over a distance of the tile thickness (10 mm) will not dominate the error budget.

%We have developed two independent methods for mapping the refractive indices~\cite{allison-2}, one based on deflections of a laser beam traversing the tiles and the other using a beam of relativistic electrons from a linear accelerator (linac).
%This article concerns the latter.

This paper describes our development of a system to measure the refractive indices using a beam of relativistic electrons from a linear accelerator.

\begin{figure}[h]
\vspace*{0.0cm}
\centerline{\includegraphics[width=0.8\textwidth,angle=270.]{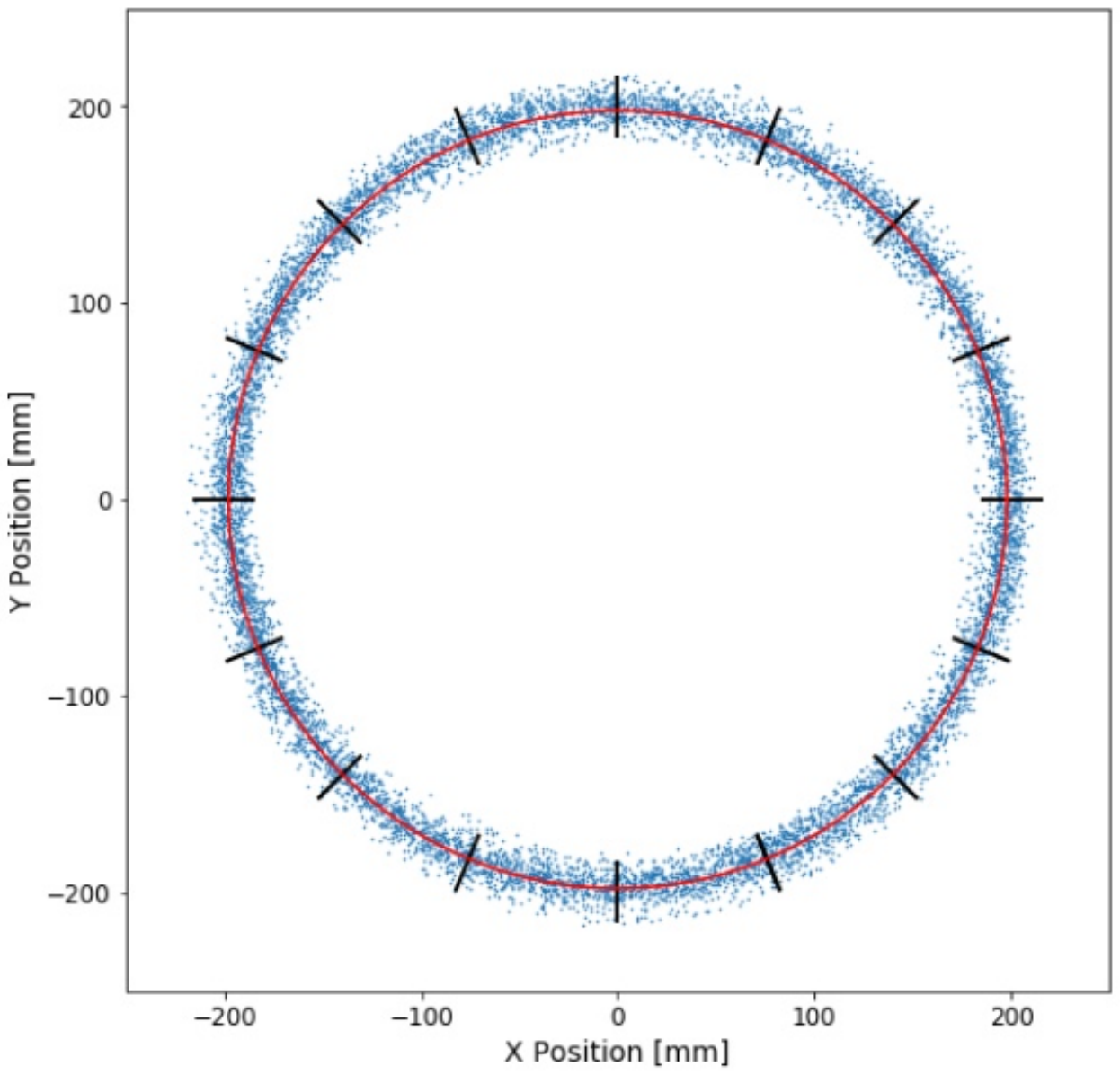}}
\vspace*{0.0cm}
\caption{
An illustration of the calibration concept.
An aerogel tile placed upstream of a detector plane will give rise to a ring of
photon impacts, shown as blue dots, when traversed by a normally-incident electron beam. 
A set of linear CCDs, shown as black radial lines, can sample the photon 
distribution and a circle, shown in red, can be fitted to the data.
}
\label{concept}
\end{figure}

\section{Calibration Setup}\label{setup-section}

The electron-beam calibration concept is illustrated in Fig.~\ref{concept}.
A normally-incident beam of electrons traversing a tile will result in a ring of photon impacts on a detector plane downstream of the tile. 
Measurement of these points will allow a circle to be fitted and the radius of the circle can be used in a calculation of the refractive index of the region of the tile traversed by the electrons. 
The width of the ring results primarily from multiple scattering of electrons in the tile and to a lesser extent from geometric aberration caused by the thickness of the tile, by chromatic dispersion effects, and by the width and divergence of the electron beam.

\begin{figure}[h]
\vspace{0cm}
\centerline{\includegraphics[width=0.8\textwidth,angle=0.]{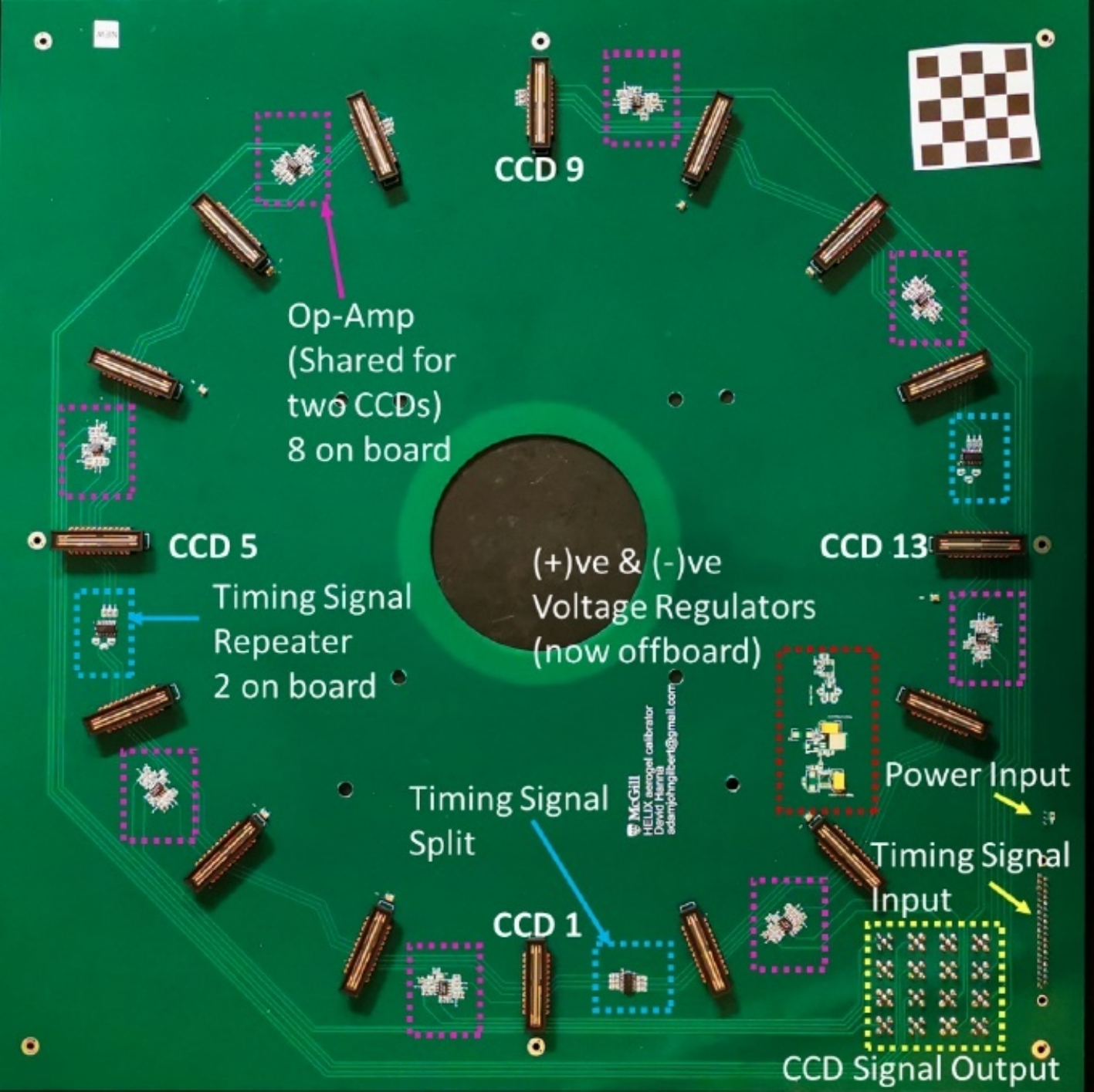}}
\vspace{0cm}
\caption{
A photograph of the one-dimensional CCD arrays on their circuit board, as viewed from the position of the aerogel tile.
They are oriented radially, with their centres on a circle with radius 200 mm. 
The checkerboard pattern, with 10 mm squares, indicates the scale.
}
\label{CCD-board}
\end{figure}

The most efficient way to measure the ring would be to use the HELIX RICH detector plane, as it would sample a larger fraction of the ring, but this was not an option at the time the tiles were being calibrated.
Instead, we opted to use an array of linear CCDs, shown schematically as radial black lines in the figure, to sample the photon distribution, and rely on a relatively intense beam to provide the necessary statistics.
We use 16 Toshiba TCD1304DG linear CCDs~\cite{Toshiba}, each of which comprises 3648 active pixels, each 200 $\mu$m wide and 8 $\mu$m long. 
Spectral response peaks near 550 nm and is lower by 20\% at 400 nm where a hard cut-off due to the glass window material that covers the CCD elements occurs. 
The CCD arrrays are located in sockets on a 500 mm $\times$ 500 mm printed circuit board, oriented radially with their midpoints located on a circle of radius 200 mm and spaced at azimuthal intervals of 22.5$^\circ$, as shown in Fig.~\ref{CCD-board}.

\begin{figure}[h]
\vspace{0cm}
\centerline{\includegraphics[width=0.8\textwidth,angle=0.]{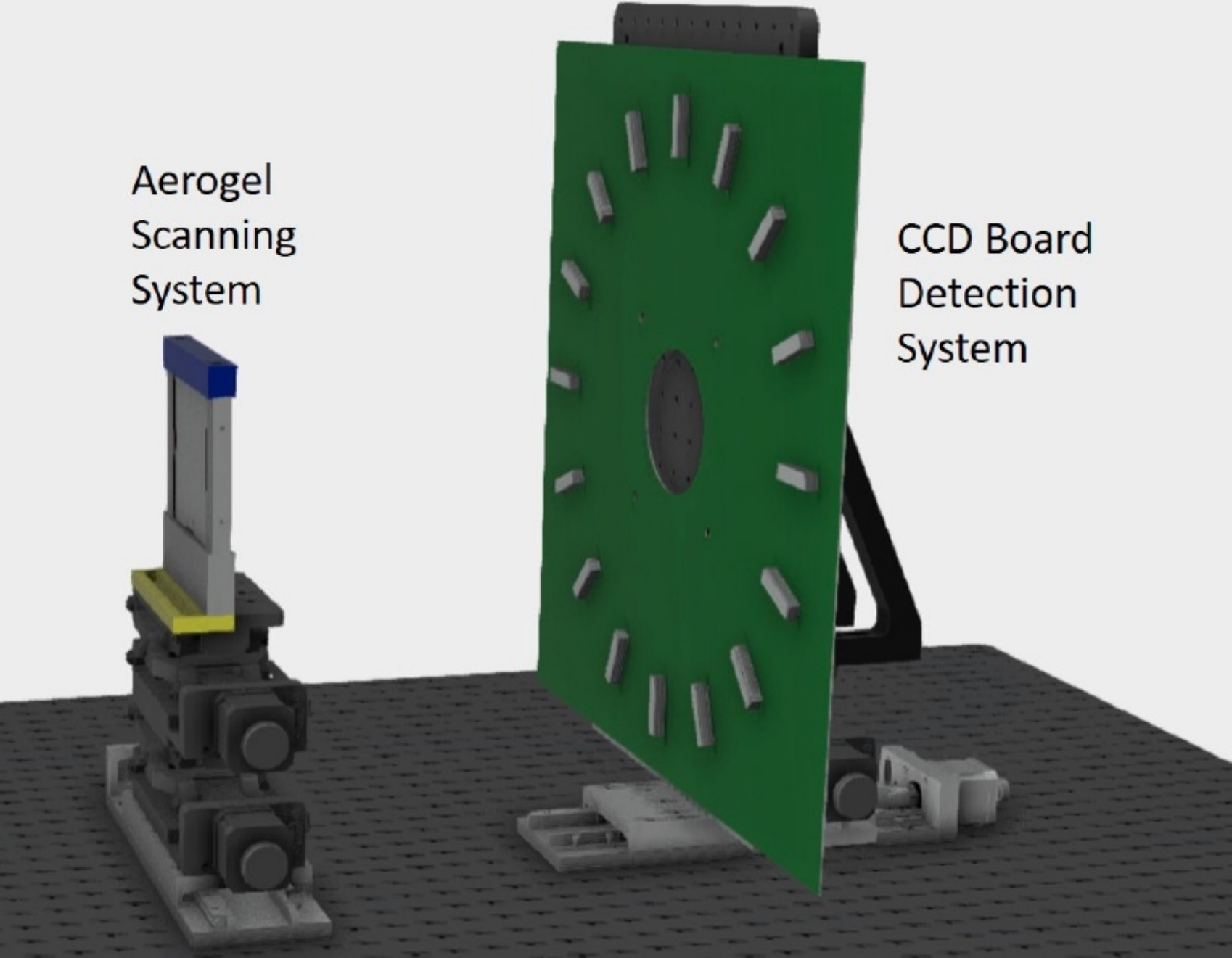}}
\vspace{0cm}
\caption{
A rendering of the calibration setup.
The aerogel tile, mounted on an x-y translation device is shown at the left such that the beam, which enters from the left, is normally incident on it.
The CCD board is located approximately 280 mm downstream of the tile and is mounted on translator stages that enable one to place the maximum of the ring distribution near the centre of each CCD. 
}
\label{setup}
\end{figure}

A rendering of the scanning setup is shown in Fig.~\ref{setup}.
The aerogel tile is held in a frame bolted to the top of a stack of translation stages, one of which is used for horizontal (x) displacements over a range of 100 mm.  
Two more stages, each with a range of 50 mm move the tile vertically (y).
With these one can scan the tile with respect to the electron beam, which defines the z axis of the coordinate system.
The tile is mounted such that the beam enters the face that will be uppermost in the HELIX payload, i.e. the electrons pass through the tile in the same direction as will the cosmic rays.

Downstream of the tile, the CCD board is mounted on a stack of three stages that allow translations in x, y, and z. 
These are used to make fine adjustments before a scan to optimize the light distributions in the CCDs such that the maxima are approximately centred. 
These adjustments are small and infrequent and are never changed during the course of the scan of a given tile.
Adjustments in the z direction affect the expansion distance for the Cherenkov cone and are therefore corrected for in the data analysis.
See section \ref{protocol} for details.

\section{Electron Beam}

The beam used for calibration was provided by the Vickers electron linac at the National Research Council in Ottawa~\cite{ross}.
The electron energy distribution is Gaussian with mean $\mu_E \simeq~$35 MeV and standard deviation $\sigma_E \simeq~$0.4\%. 
The beam profile can be described by a two-dimensional Gaussian with widths $\sigma_{x,y} \simeq 2$ mm, as measured by a profile monitor 110 mm upstream of the front face of the tile being measured.
The beam divergence was not measured during this work. It was used as a free parameter in simulation studies and the results were consistent with a published measurement~\cite{ross}.

The linac current was set to 90 nA and was delivered as 2.5 $\mu$s pulses at a rate of 60 Hz, with each pulse containing approximately $10^{10}$ electrons.
These settings resulted in a signal in the CCDs that was well above thermal noise but far from saturation.

\section{Electronics}

The exposure times and readout of the CCDs are controlled by signals generated by NIM and CAMAC modules located in nearby crates and sent to the detector board via a ribbon cable.
Coaxial 50-ohm cables, one per CCD, are used to transport the CCD signals to the digitizing electronics.
These electronics comprise 16 channels of Acqiris U1063A DC270 1 GS/s 8-bit digitizers, configured as four 4-channel modules in a dedicated crate controlled by a single-board ADLINK cPCI-6620 computer housed in the same crate.  

\subsection{CCD Control}

The CCDs require three signals. The shift gate (SH) signal is a pair of pulses, each 2 $\mu$s wide, that define the exposure time for the CCD, set to 20 $\mu$s.
The readout of the data from the CCD pixels is started by the integration clear gate (ICG), a pulse with width 5 $\mu$s that starts 0.5 $\mu$s before the second SH pulse.
Charges from the CCD pixels are clocked out by the master clock signal ($\phi$M), a 2 MHz square wave.
For details see~\cite{Toshiba}.

The CCD control pulses are generated on receipt of a trigger from the linac control, sent in advance of the pulse of accelerated electrons.
There is considerable jitter between this trigger and the arrival of the electrons, which is the motivation for the 20 $\mu$s exposure of the CCDs even though the beam pulse is only 2.5 $\mu$s in length.
The apparatus is covered with a light-tight box during data-acquisition runs so the long exposure does not increase background significantly.

\subsection{Digitizers}

The data from all CCDs are digitized simultaneously using the 16 FADC channels, triggered by the ICG pulse.
The internal timing of the FADCs is set to match the arrival times of data clocked from the CCDs (one every 2 $\mu$s with $\phi$M = 2 MHz) so one 8-bit FADC value is produced for each pixel.
A readout cycle takes 7.4 ms for 3648 active and 46 dummy pixels.
This is short enough to fit between linac pulses; they occur every 16.7 ms since the linac runs at 60 Hz.

\begin{figure}[h!]
\vspace{0cm}
\centerline{\includegraphics[width=0.8\textwidth,angle=0.]{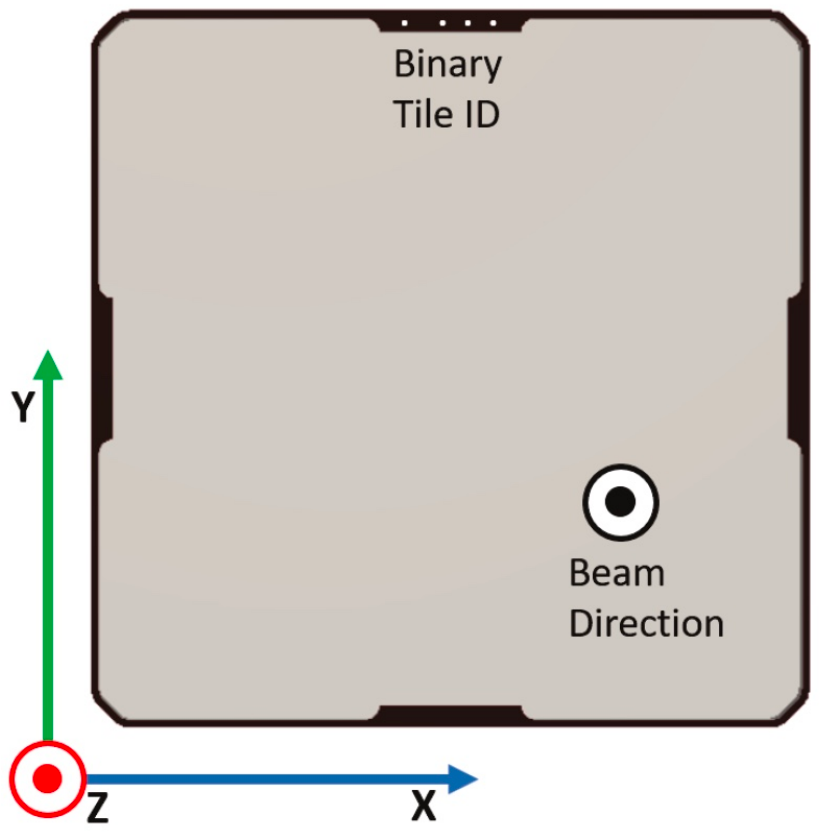}}
\vspace{-2cm}
\caption{
A schematic of an aerogel tile, illustrating the coordinate system used in the scans.
Each tile is enclosed in an aluminum frame, seen as the black structure at the 
edges. 
(The four tabs are to prevent the tile from shaking loose during flight or transport and a unique binary identifier is etched on one of them.)
}
\label{coord_sys}
\end{figure}

\section{Scanning Protocol}\label{protocol}

Each tile is scanned following the same procedure. 

\begin{itemize}

\item{The beam is disabled and a new tile is installed after the one scanned in the previous run is removed.}

\item{A `dark run' of 100 readout cycles is performed, to provide pedestal values to be used later.}

\item{The beam is re-enabled and some data are acquired to confirm that the CCD distributions are approximately centred by utilizing an on-line ‘quick-look’ diagnostic program.
Adjustments to the position of the CCD board (as described in section~\ref{setup-section} ) may be made at this time.}

\item{The tile is positioned such that the beam passes through the point $x$ = 5 mm, $y$ = 95 mm (see Fig.~\ref{coord_sys} for the coordinate system) and a run of 100 beam pulses, each read out separately, is performed.}

\item{The tile is repositioned so that the beam passes through $x$ = 10 mm, y = 95 mm and another run is performed.}

\item{The procedure is repeated until the entire grid of 19 $\times$ 19 points has been covered. 
The scan covers a grid with 5 mm pitch, starting and ending 5 mm from the tile edges.}

\end{itemize}

The data are buffered for the 100 pulses at each scan point and are written to disk in a custom binary format 
while the tile is moved to the next scan position.
The binary format enables fast file writing and easy access by the `quick-look' program. 
They are converted to Flexible Image Transport System (FITS) files offline.

Data acquisition and file writing take approximately 3.3 s per scan point and movement between
scan points takes approximately 2.5 s. 
An entire 361-point runs takes approximately 35 minutes.
Changing to a new tile requires 15 minutes.

All told, 52 tiles were scanned following this procedure and several were measured more than once, for reproducibility studies.

\section{Data Analysis}

The data set comprises a FITS file for each scan point on each tile.
Each file contains 100 `images', one per beam pulse, where each image contains the 16 data arrays for the CCDs.
Each array is 3694 pixels long, corresponding to the 3648 active and 46 dummy pixels in the CCD.
In the following we use pixels 51 through 3650. 

\begin{figure}[h]
\begin{tabular}{cc}
\begin{minipage}{0.45\linewidth}
\centerline{\includegraphics[width=1.1\textwidth,angle=0.]{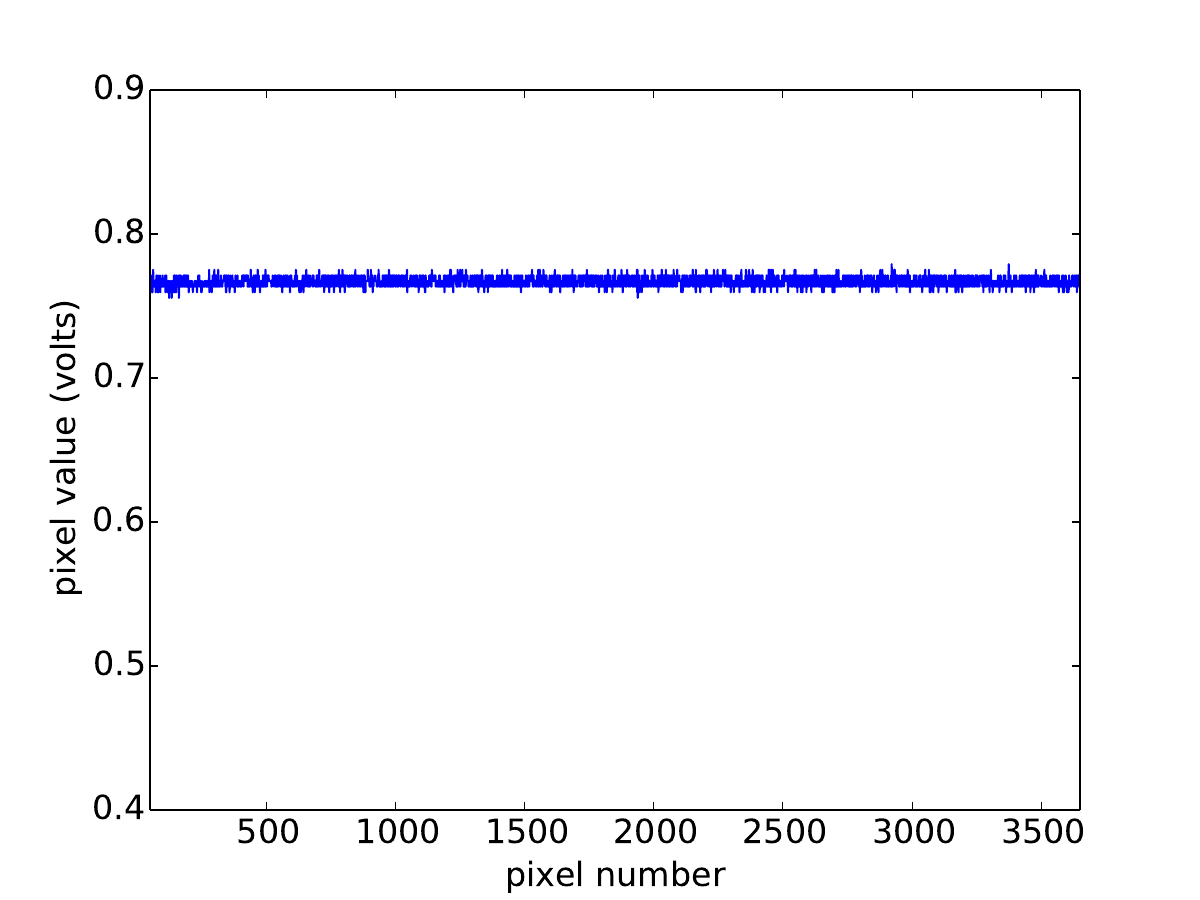}}
\vspace{0mm}
\end{minipage}

&

\begin{minipage}{0.45\linewidth}

\centerline{\includegraphics[width=1.1\textwidth,angle=0.]{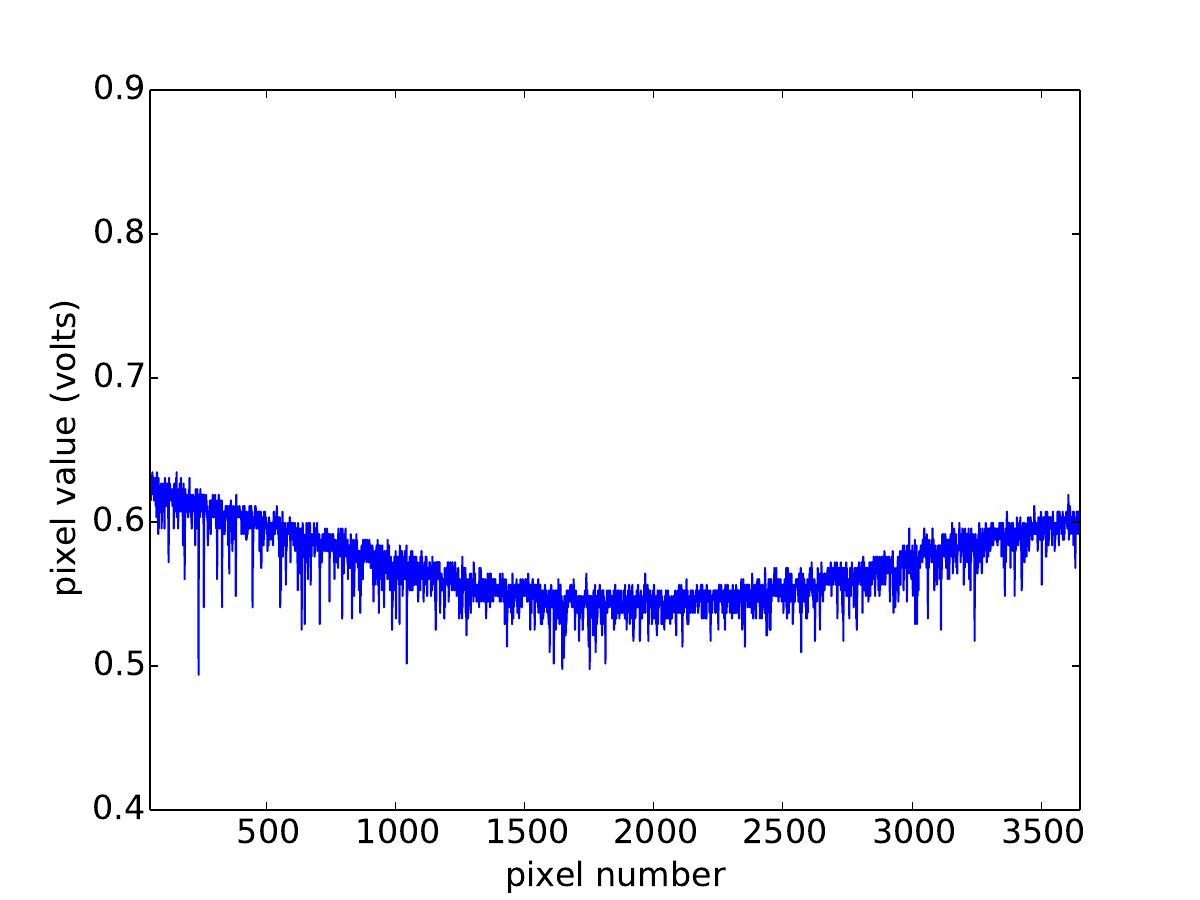}}
\vspace{0mm}
\end{minipage}
\end{tabular}
\caption{
Left: A sample dark data array from one of the CCDs. Pixel readings 51 through 3650 are plotted. 
Right: A sample active data array made with data taken after a pulse of electrons had traversed the aerogel tile.
}
\label{trace_1_2}
\end{figure}

A sample array from a dark run and a sample array from a data run are shown in Fig.~\ref{trace_1_2}.
It is clear from these plots that the raw data are `inverted', with the baseline at a larger voltage value than the signal.
An estimator for the charge in a given pixel is obtained by subtracting the pixel value in a data run from its value from the dark run.

\begin{figure}[h]
\begin{tabular}{cc}
\begin{minipage}{0.45\linewidth}
\centerline{\includegraphics[width=1.1\textwidth,angle=0.]{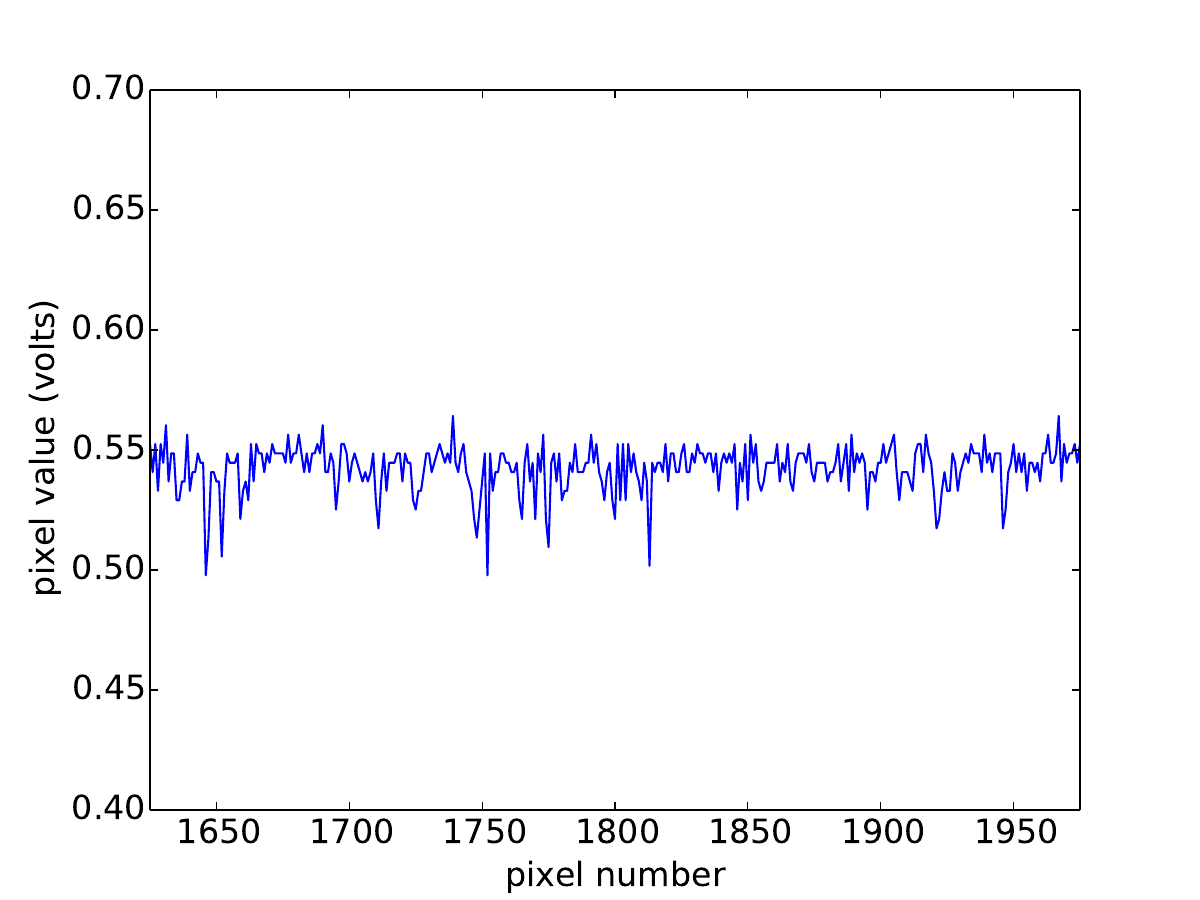}}
\vspace{0mm}
\end{minipage}

&

\begin{minipage}{0.45\linewidth}

\centerline{\includegraphics[width=1.1\textwidth,angle=0.]{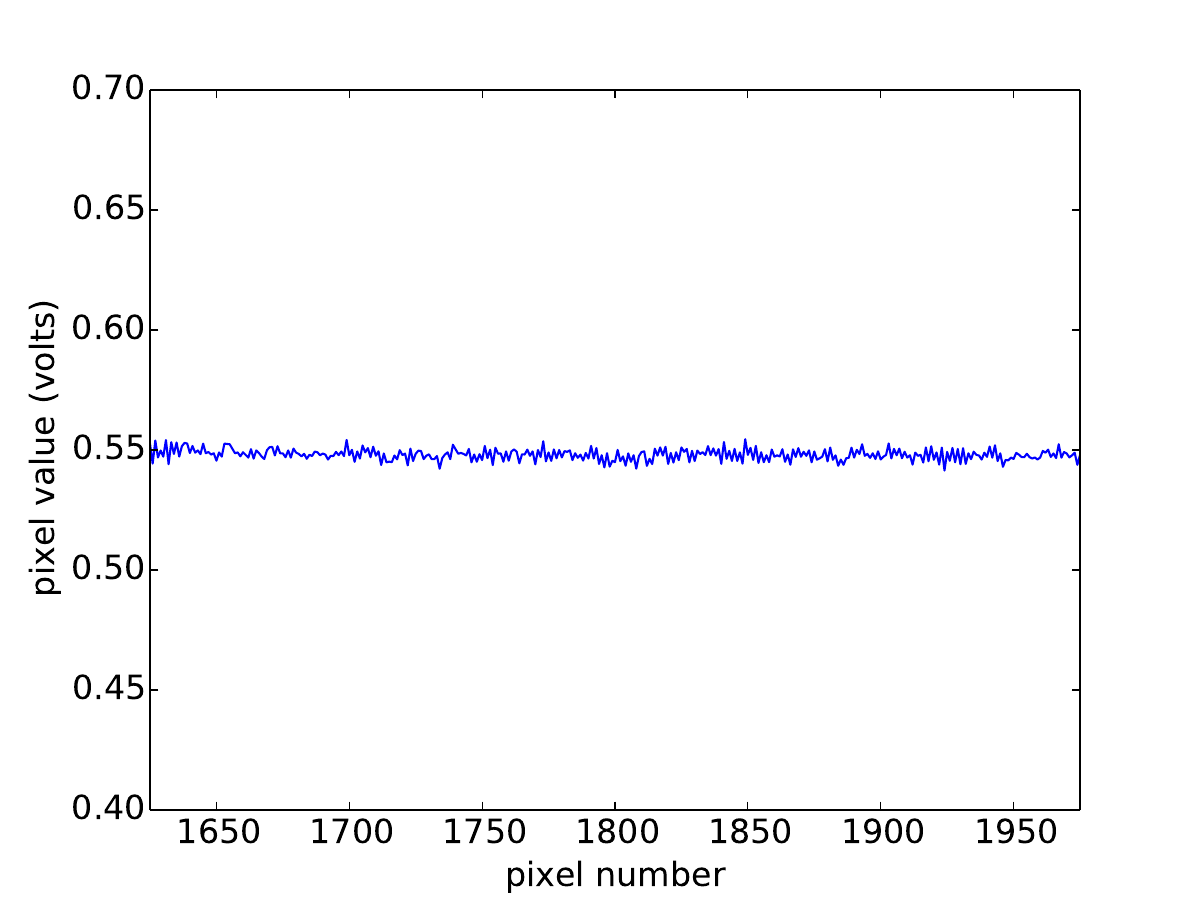}}
\vspace{0mm}
\end{minipage}
\end{tabular}
\caption{
Left: readings from 350 pixels near the centre of a single CCD for a single image, showing large pixel-to-pixel fluctuations.
Right: readings from the same pixels averaged over 100 images showing that the fluctuations average out and are not due to any systematic effects.
}
\label{raw_1_2}
\end{figure}

One notices large pixel-to-pixel fluctuations in Fig.~\ref{trace_1_2} (right).
These are not indicative of anything more than statistical fluctuations.
A plot of the values from the central 350 pixels of a CCD for an arbitrarily chosen image is shown in the left panel of Fig.~\ref{raw_1_2}; one sees the fluctuations more clearly. 
In the right panel of the same figure the 100-image averages of the same pixels are plotted. 
The pixel-to-pixel variations are considerably reduced, showing that pixel-dependent
gains or efficiencies are small and treating all pixels in the same way is a practical way to proceed. 
(Note that without `flat-fielding' the CCDs, there will be small pixel-dependent differences which can affect the precise value of the $\chi^2$ of certains fits. 
See the discussion in section~\ref{ring-params}.)

\section{Computation of Ring Parameters}\label{ring-params}

The first step in computing the radius of each Cherenkov ring is to find the maximum of the distribution of charge in each CCD.
As can be seen from Fig.~\ref{trace_1_2}, the distribution of charge in the CCDs is broad and does not go to zero within the limits of the CCDs.
This is mostly due to multiple scattering of the electrons in the aerogel tile, a big effect for 35 MeV electrons; simple Moliere theory predicts a broadening of about 20 mm (standard deviation) of the ring at the CCD. 
Other effects are discussed in the section on simulations.
It means that using the weighted mean of the distribution is not helpful and we are motivated to use the coordinate of the pixel at which the distribution reaches its maximum as our estimator for the position of the Cherenkov ring.
This is analogous to using the most probable value in a distribution of charge samples when making dE/dx measurements in particle physics.

\begin{figure}[h]
\vspace{0cm}
\centerline{\includegraphics[width=0.9\textwidth,angle=0.]{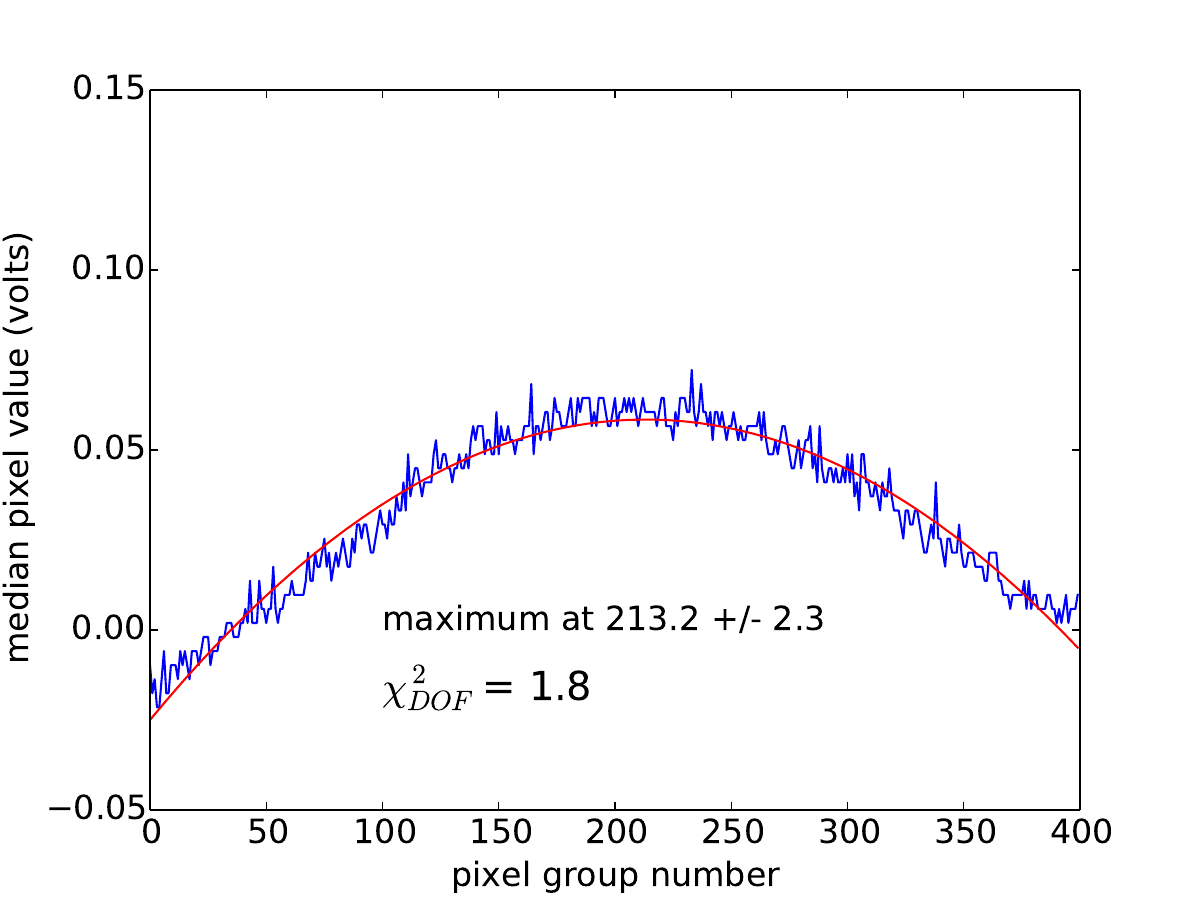}}
\vspace{0cm}
\caption{
A example of the parabolic fit used to find the maximum of the CCD distribution.
The data points have been pre-processed as described in the text and the baseline adjusted by subtracting the average of the first and last points.
}
\label{para_1}
\end{figure}

To compute the maximum point of each CCD charge distribution we fit a parabola to the readings and use the fit parameters to calculate the point at which the maximum occurs.
Prior to the fit, the 3600 readings from pixels 51 through 3650 are compressed to 400 by taking medians of consecutive groups of 9. 
This reduces crowding in plots and eliminates the effects of any outliers.
The loss of positional resolution is inconsequential, given the broad distribution of the Cherenkov light compared with the CCD pixel size.
The pixel value at which the fit's maximum occurs, along with its uncertainty and the $\chi^2_{DOF}$ of the fit, are saved in a file, to be used in the next analysis stage.
An example plot of the compressed data and corresponding parabola fit is shown in  Fig.~\ref{para_1}.
The point uncertainties used in the fit are estimated by comparing the value of the point to the average of its two nearest neighbours. 
This assumes a smooth, linear change in the point values. 
The resulting $\chi^2_{DOF}$ is large, indicating that the uncertainties are too small and/or the hypothesis of a parabolic shape is incorrect.
Both explanations are likely; there can be non-statistical fluctuations from one CCD pixel to another caused by differences in pixel size, for example.
These would propagate into the final value of each fitted point.
Likewise, the use of a parabolic fit is not motivated by a physical model of the ring profile but is a use of a limited number of terms in a Taylor expansion.
Fitting over a smaller range will result in a smaller $\chi^2_{DOF}$ but will reduce the utility of this number for excluding corrupted images, as described below.

The 16 maxima are fitted to a circle in the next stage of the analysis.
Before the fit is made, the values of the maxima in pixel numbers on the CCD chips are converted to mm values on the CCD board.  
The circle fit has three parameters, $x_c$, $y_c$ and $r$,  with the first two being the coordinates of the centre and the last being the radius. 
The centre is only of interest as a check for pathologies; the radius is the important parameter.

\begin{figure}[h]
\vspace{0cm}
\centerline{\includegraphics[width=0.8\textwidth,angle=0.]{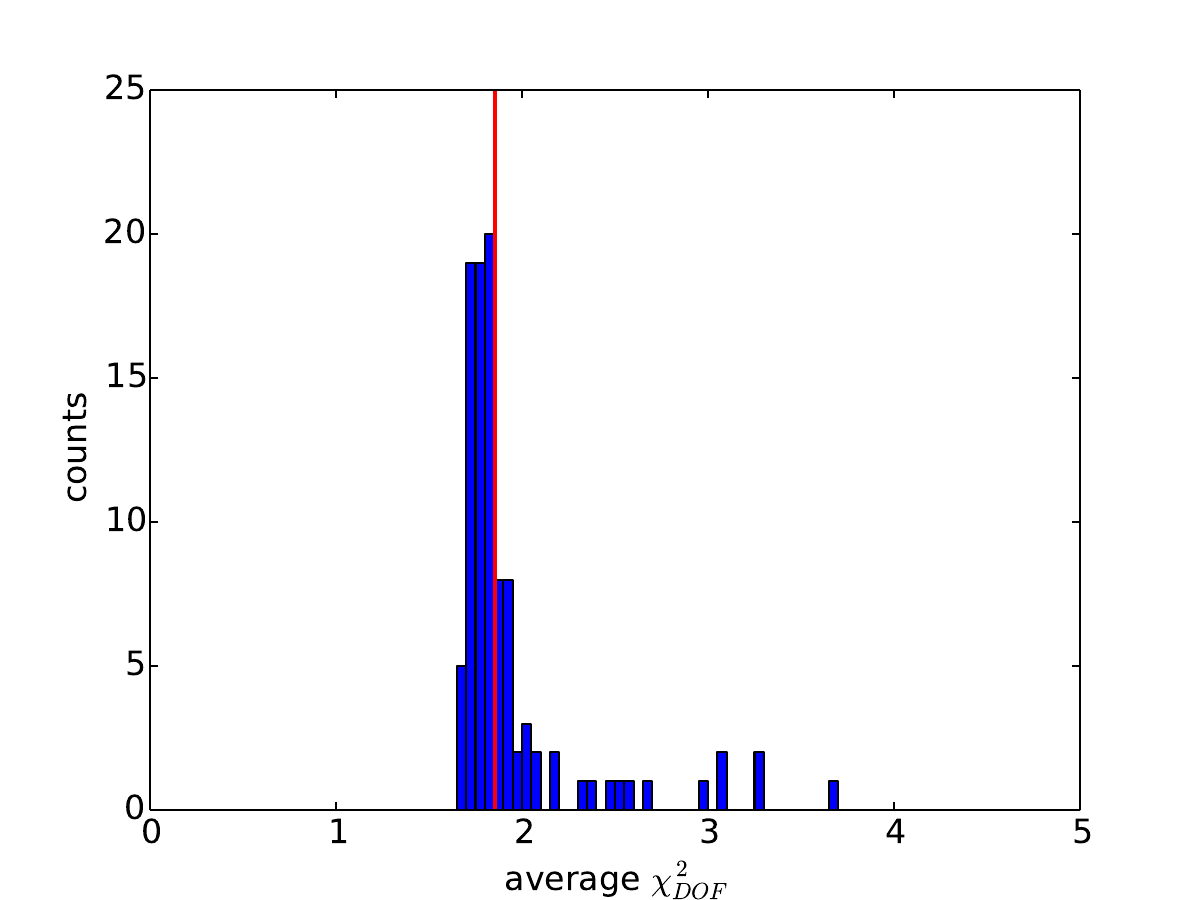}}
\vspace{0cm}
\caption{
Histogram of the $\chi^2_{DOF}$ parameters, averaged over all 16 channels for each of the 100 beam pulses, from fits like those shown in Fig.~\ref{para_1}.
The values do not peak near 1.0 due to uncorrected systematic errors and/or because the parabolic fit is not a perfect description of the data.
Nevertheless, all distributions display a clustering at lower values and a tail to higher values. 
A cut that includes the lowest 65\% of the entries, shown here as a red line, is applied to the data to 
eliminate events with corrupted data.
}
\label{chi_avg}
\end{figure}

Before performing the circle fit we make a cut on the $\chi^2_{DOF}$ values from the parabola fits. 
The cut is to eliminate images where all 16 readout channels were affected by a coherent background effect caused by intermittent noise in the circuitry.
The effect was discovered as a distortion in approximately one third of the flat data arrays recorded in dark 
runs but is also seen in a similar fraction of beam-on arrays.
It is easy to eliminate by ranking images according to the 16-channel-average $\chi^2_{DOF}$ and using the lowest 65\%.
A sample distribution of $\chi^2_{DOF}$ values for a typical image (100 beam events) is shown in Fig.~\ref{chi_avg}.

\begin{figure}[h]
\vspace{0cm}
\centerline{\includegraphics[width=1.1\textwidth,angle=0.]{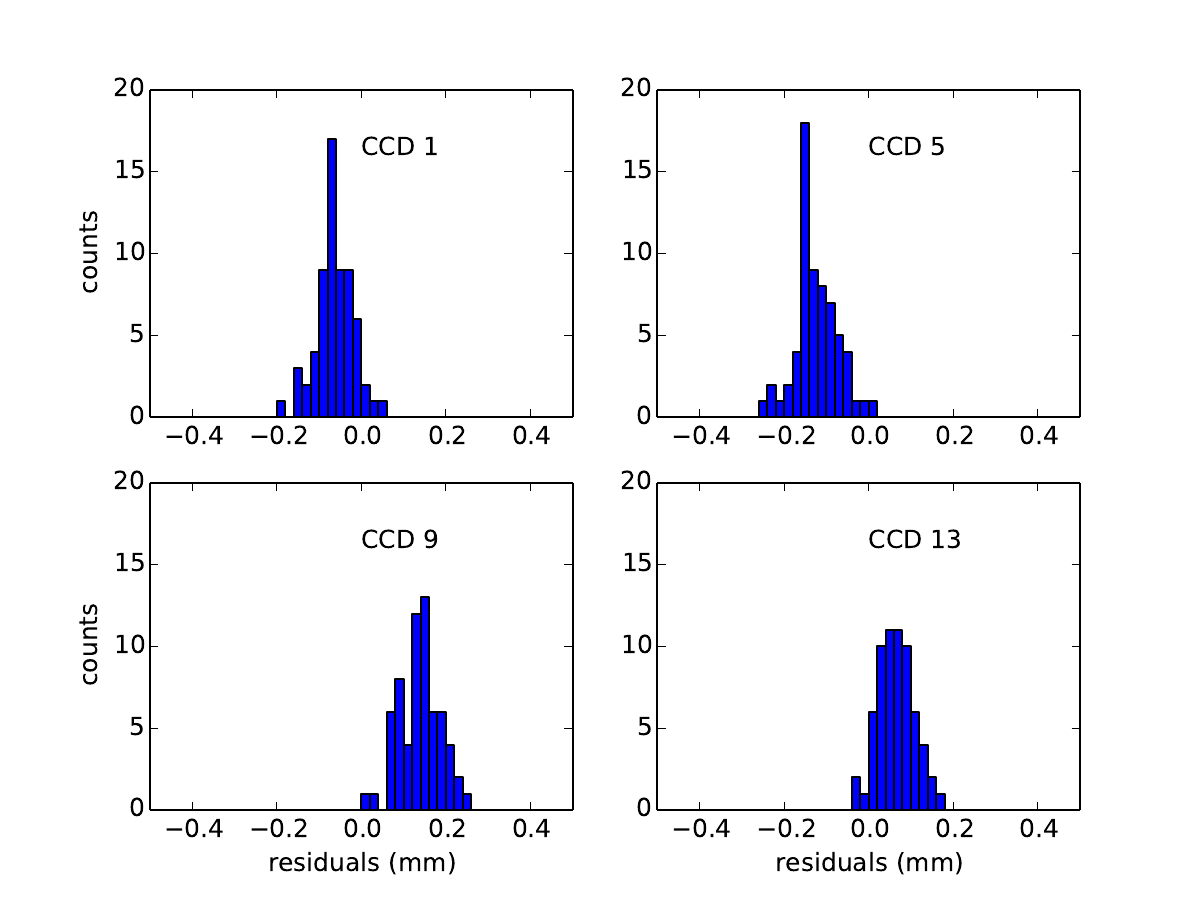}}
\vspace{0cm}
\caption{
Sample plots of the residuals (difference between the radius estimate using the CCD information and the radius given by the fit of a circle to all 16 estimates) for a single scan point on an arbitrary tile.
Each plot corresponds to a separate CCD. 
The distributions have similar widths but different mean values.
The mean values, all within 0.3 mm of 0.0, are probably due to chip-placement tolerances on the CCD board.
}
\label{resid}
\end{figure}

Sample residual distributions from the circle fits for a single scan point on an arbitrary tile are plotted in Fig.~\ref{resid}. 
Only four histograms are plotted to save space; the other 12 look very similar.
The widths are all approximately the same ($\sigma \simeq 0.1$ mm). 
The means are significantly different from each other but are all within about 0.3 mm of 0.0. 
Moreover, the means do not change as one scans over the tile, or if one scans a different tile. 
We thus attribute them to tolerances on the placement of the CCD chips;  they can be accounted for with an ad-hoc correction to the CCD positions.

\section{Radii and Refractive Index}
\label{sec:radii}
A complete scan of an aerogel tile results in 19 $\times$ 19 files to process but the edge scan points are of lower quality for a number of reasons.
These include small cracks or chips caused by the water-jet cutting and effects of the Cherenkov cone being partly blocked by the aluminum frame surrounding the tile. 
Thus we restrict further analysis to the 289 (17 $\times$ 17) points running  
from 10 mm to 90 mm, in 5 mm steps, in both the $x$ and $y$ directions.

\begin{figure}[h]
\vspace{0cm}
\centerline{\includegraphics[width=1.1\textwidth,angle=0.]{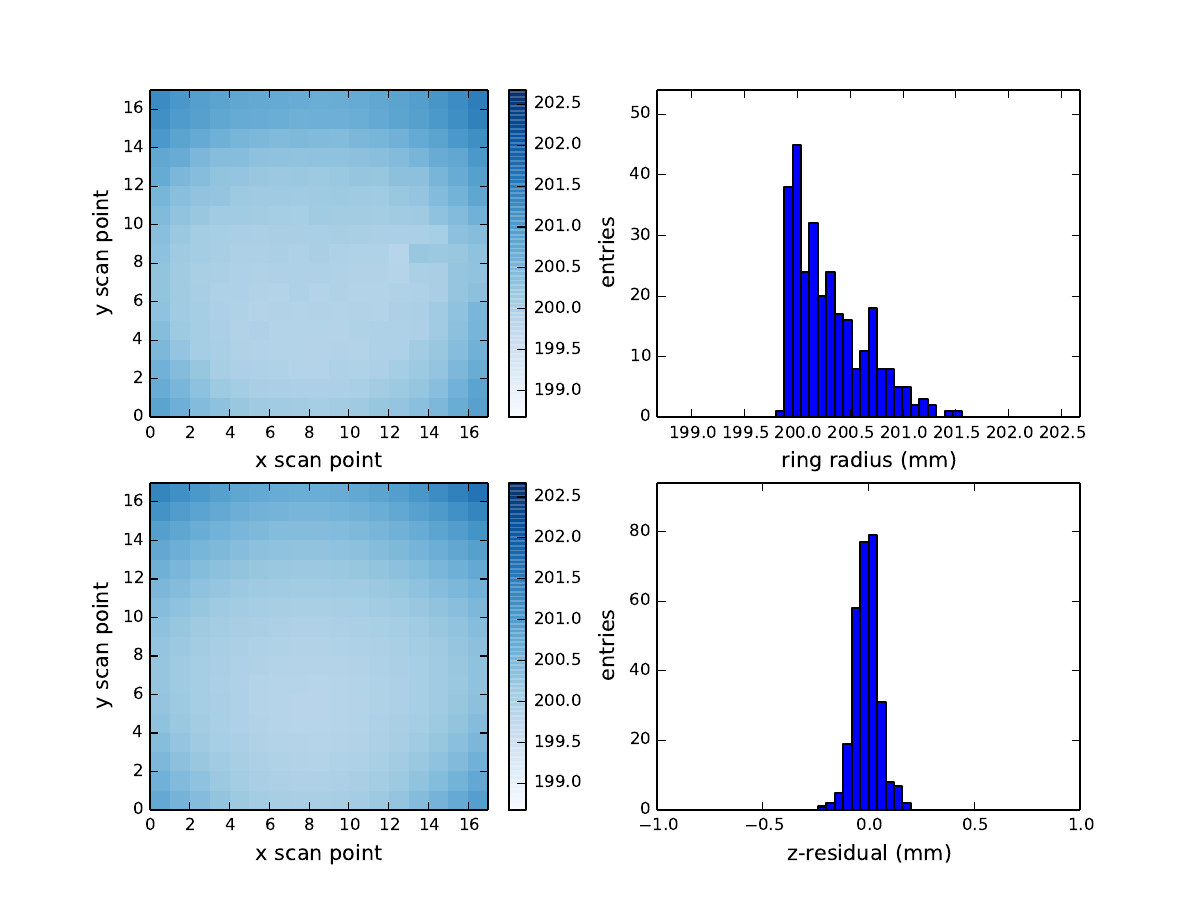}}
\vspace{0cm}
\caption{
Results from the circle fits for a scan of a single aerogel tile.
Upper left: the radii for the 289 (17 $\times$ 17) scan positions.
Upper right: the radii values, histogrammed.
Lower left: values from a two-dimensional (9 parameter) parabolic fit to the data.
Lower right: residuals from the fit, histogrammed.
}
\label{map_1}
\end{figure}

\begin{figure}[h]
\centerline{\includegraphics[width=1.1\textwidth,angle=0.]{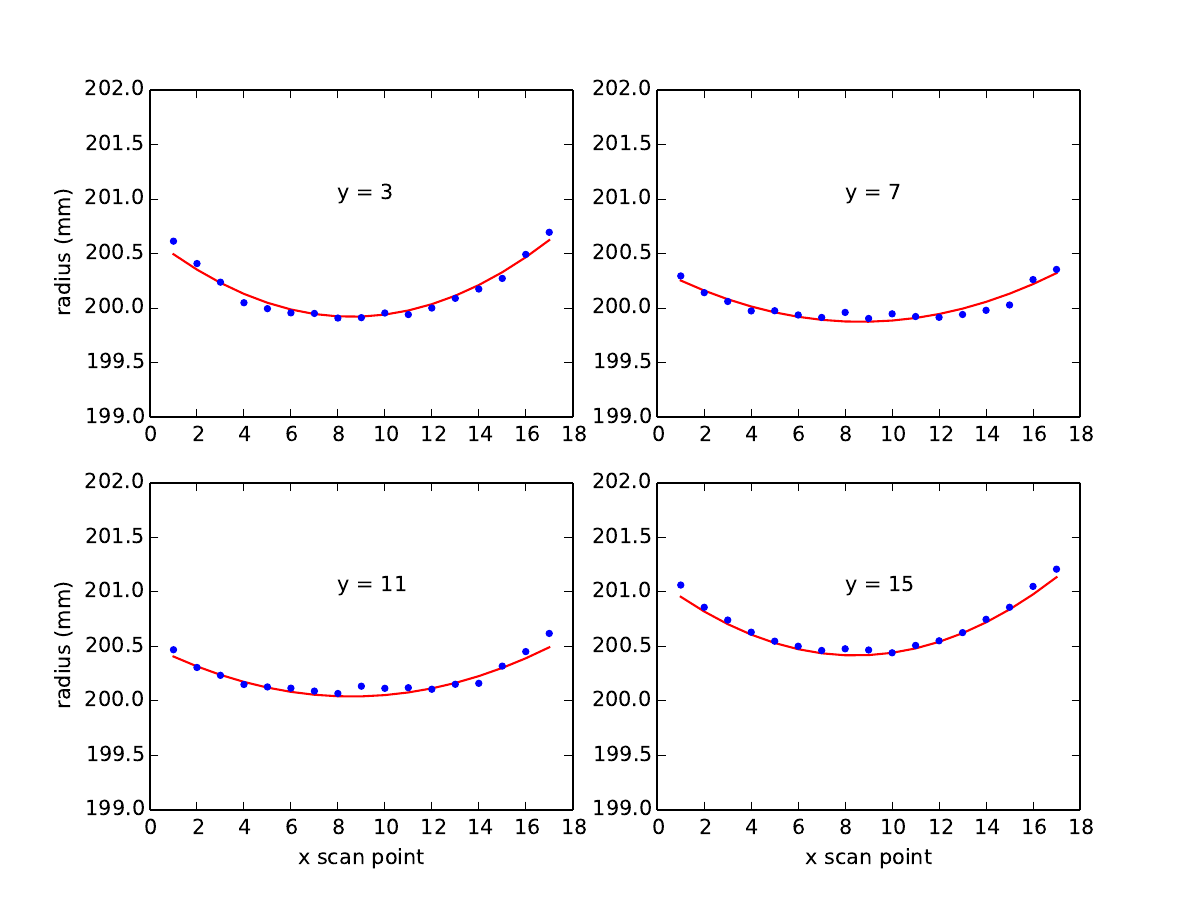}}
\vspace{0mm}
\caption{
Sample comparisons of the measured data shown in Fig.~\ref{map_1} with the values given by a global, nine-parameter
two-dimensional parabolic fit.
The x dependence is plotted for four y slices; others are similar.
The abscissas are in units of scan steps (5 mm) and the ordinates are the ring radii in mm.
The y values are also in units of 5-mm scan steps.
}
\label{map_2}
\end{figure}

The radii are directly related to the Cherenkov angle and thereby to the local refractive index.
We begin by looking at the radii for a given tile scan.
An example of the radii's distribution across a 17 $\times$ 17 position scan is shown in Fig.~\ref{map_1} (upper left panel).
Also shown is a histogram of the values (upper right panel).
These points are well fitted by a two-dimensional (nine parameter) parabola.
The parabola's predictions for the scan points are shown in the lower left panel and the residuals from the fit are histogrammed in the lower right panel.
Line-by-line plots of the points and fit are shown in Fig.~\ref{map_2}.
The fit is a remarkably good description of the data.

To relate the ring radius, $r$, to the refractive index, $n_a$, in the aerogel tile we need to account for the refraction that takes place at the surface of the aerogel tile.
The light emitted inside the tile at angle 
$\theta_c$ propagates between the tile surface and the detector plane, a distance $d$ downstream, at a different angle, $\theta$.
From Snell's law we calculate $n_0 {\rm sin} \theta = n_a {\rm sin}\theta_c $ where $n_a$ is the refractive index of aerogel and $n_0$ is the index of air, about 1.0003 for blue light.
This leads to the expression for tan$\theta_c$, which needs to be solved numerically,
% Making "equation" for later referencing
\begin{equation}
    {\rm tan}\theta_c = n_0 \beta (r - z_e {\rm tan}\theta_c)/\sqrt{(r - z_e {\rm tan} \theta_c)^2 + d^2},
    \label{eqn_thetac}
\end{equation} 
where $z_e$ is the distance between the point where the Cherenkov emission occurs and the exit face of the tile. 
The emission point varies continuously between the two faces of the tile so we make the approximation that $z_e$ is constant and has the value of half the tile thickness at the point being scanned.
$\beta$ is 1.0 at the beam energy used.

\begin{figure}[h]
\vspace{0cm}
\centerline{\includegraphics[width=1.1\textwidth,angle=0.]{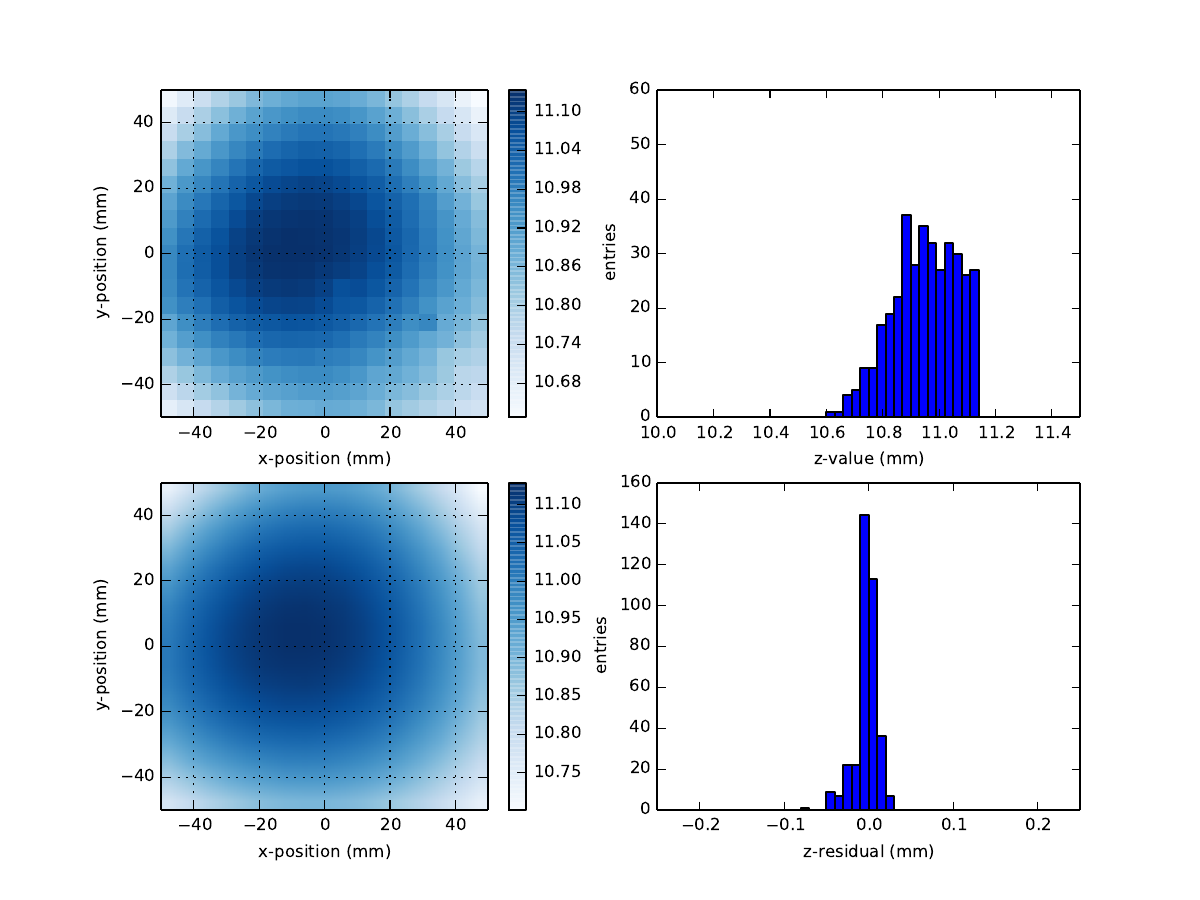}}
\vspace{0cm}
\caption{
Results from a scan of a typical aerogel tile using a Mitutoyo QV606 CMM at TRIUMF. 
In the upper left panel the height in mm of the aerogel surface above a reference plane is plotted for each position on a 19 $\times$ 19 grid with 5 mm pitch. 
The values are histogrammed in the upper right panel.
The lower left panel displays a heat map of the values given by a two-dimensional parabolic (nine parameters) fit to a 17 $\times$ 17 subset of the data that excludes edge points.
Residuals for the fit are shown in the histogram in the lower right panel.
}
\label{tile_31_1}
\end{figure}

\begin{figure}[h]
\vspace{0cm}
\centerline{\includegraphics[width=1.1\textwidth,angle=0.]{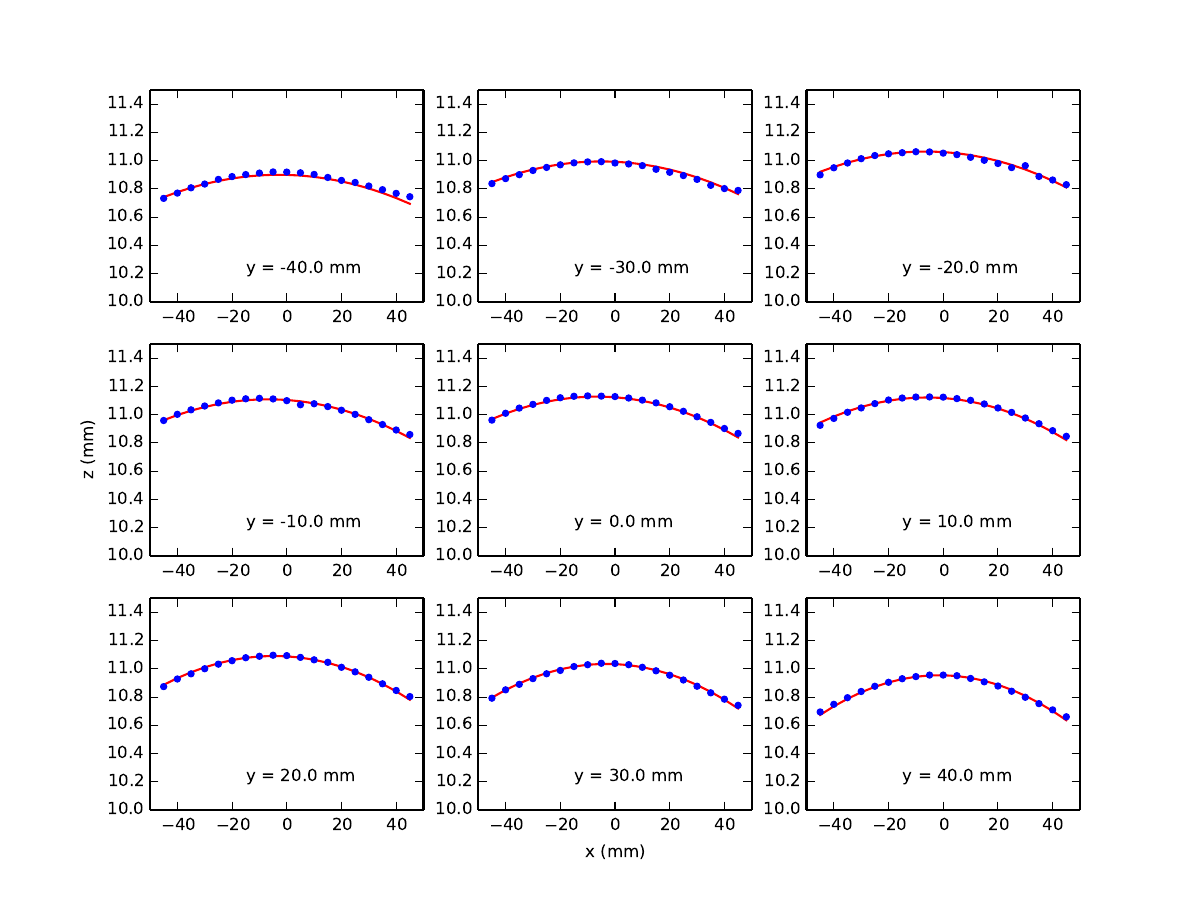}}
\vspace{0cm}
\caption{
Slices in x, for a series of y values, showing the ability of the nine-parameter fit (red) to describe the data (blue points) shown in Fig.~\ref{tile_31_1}.
This illustrates the smooth behaviour of the tile surface and how well it can be parameterized by a simple function.
}
\label{tile_31_2}
\end{figure}

An additional complication arises because the tiles are not flat; there are centre-to-edge differences of up to 0.5 mm. 
These change the local value of $d$, so we can expect variations in $r$ of 0.3 to 0.4 mm that can simply 
arise from geometry, independent of changes of the refractive index.
The tiles have all been scanned with a Mitutoyo QV606 coordinate measuring machine (CMM)
at the TRIUMF laboratory to obtain the geometrical data for implementing corrections.
The tiles all exhibit bowing; the front and back faces are roughly parallel but curve slightly in a way that is well described by a pair of two-dimensional parabolas.  
An example of the data and fit from one tile is shown in Fig.~\ref{tile_31_1} and Fig.~\ref{tile_31_2}.

\begin{figure}[h]
\vspace{0cm}
\centerline{\includegraphics[width=1.1\textwidth, angle=0]{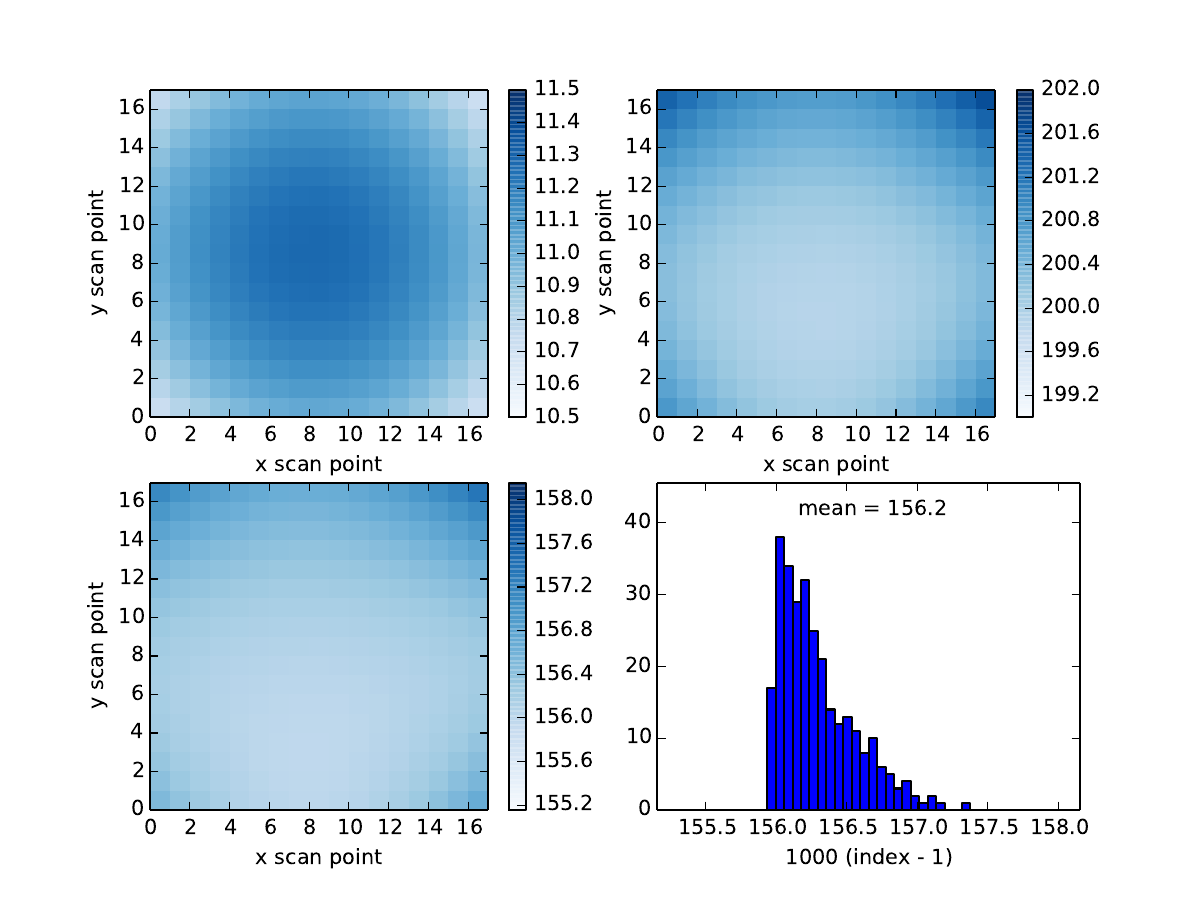}}
\vspace{0cm}
\caption{
Results from a beam scan of an aerogel tile.
The upper left panel displays the height, in mm, of the tile surface as obtained from a scan with a coordinate measuring machine with the tile, encased in its aluminum frame, laid flat on a reference plane. 
These data are used in computing the distance between the aerogel surface and the CCD detectors downstream of the tile. 
Cherenkov ring radii, in mm, for different scan positions are shown in the upper right panel.
The map of reduced refractive indices ($n' = 1000 (n-1)$), obtained using the data from the top two panels, is shown in the lower left panel.
A histogram of the values is shown in the lower right panel.
The increase of the refractive indices towards the corners of the tile is a feature common to all the tiles.
}
\label{hlx_1}
\end{figure}

The results of applying the shape corrections to data from a single tile are shown in Fig.~\ref{hlx_1}. 
From the results in Fig.~\ref{map_1} and \ref{hlx_1} we can see that point-to-point differences of 0.1 mm in ring radius or 0.1 in reduced refractive index ($n' = 1000 (n-1)$)
can be measured with this setup and procedures.
The corresponding resolution in refractive index, $\delta n/ n$, can be shown to be of order $10^{-4}$.

% Geant4 section text
%\input{geant4_section.tex}

\section{Comparison with Monte-Carlo Simulations}\label{geant}

To further understand the electron-beam data, the setup was simulated using \texttt{Geant4} (v11.0.1) \cite{AGOSTINELLI2003250, 1610988, ALLISON2016186}.
The simulation consisted of a mono-energetic (35 MeV) beam, an aerogel tile with density and composition as described by \cite{tabata-2, tabata} and a shape matching a typical tile as measured using the CMM (see Section \ref{sec:radii}).
The \texttt{QGSP\_BIC\_HP\_EMZ} `physics list' was used with \texttt{G4OpticalPhysics} enabled.  

In the full simulation, photons were generated uniformly along the electron track.
They were assigned wavelengths according to the $1/\lambda^2$ dependence of the Cherenkov effect, starting at 400 nm and weighted according to the spectral response curve of the CCDs supplied by the manufacturer~\cite{Toshiba}.
They were emitted in a cone at the angle defined by the wavelength-dependent refractive index and tracked through the rest of the tile, subject to scattering along the way, until they emerged from the tile.
The radial positions of the photons at the detector board were histogrammed with bin-widths of 8~$\mu m$ (the CCD pixel width) and the histogram contents were smoothed using a median filter with a nine-bin kernel to mimic the data-analysis methods.

Chromatic aberration is an important feature that needs to be accounted for to gain a complete understanding of the radiator in a RICH.
To study this one needs to know the wavelength dependence of the refractive index~\cite{bellunato}.
We have not measured the dependence for the tiles under study but we can use measurements of similar aerogel as a guide.
The Belle II RICH~\cite{belle-II} uses aerogel tiles made using a similar process but with a smaller average index (1.05). 
Data from those tiles have shown that $(n^2-1)$ drops by 2\% as $\lambda$ increases from 405 nm to 550 nm. 
This behaviour has been observed in similar tiles with average index near 1.12~\cite{tabata}.
We therefore simulate the wavelength dependence of the refractive index using the Sellmeier equation~\cite{sellmeier}, with parameters that produce a 2\% shift in refractive index over this wavelength range.
For the studies reported here, the refractive index at 400 nm is 1.155.

\begin{figure}
     \centering
     % \begin{subfigure}[b]{0.45\textwidth}
     %     \centering
     %     \includegraphics[width=1.1\textwidth]{figures/top-hat.pdf}
     %     \label{fig:top-hat}
     % \end{subfigure}
     % \hfill
     % \begin{subfigure}[b]{0.45\textwidth}
     %     \centering
     %     \includegraphics[width=1.1\textwidth]{figures/ste_parabola.pdf}
     %     \label{fig:parabola}
     % \end{subfigure}
     
        \vspace{-8cm}
    
        \centerline{\includegraphics[width=1.\textwidth,angle=0.]{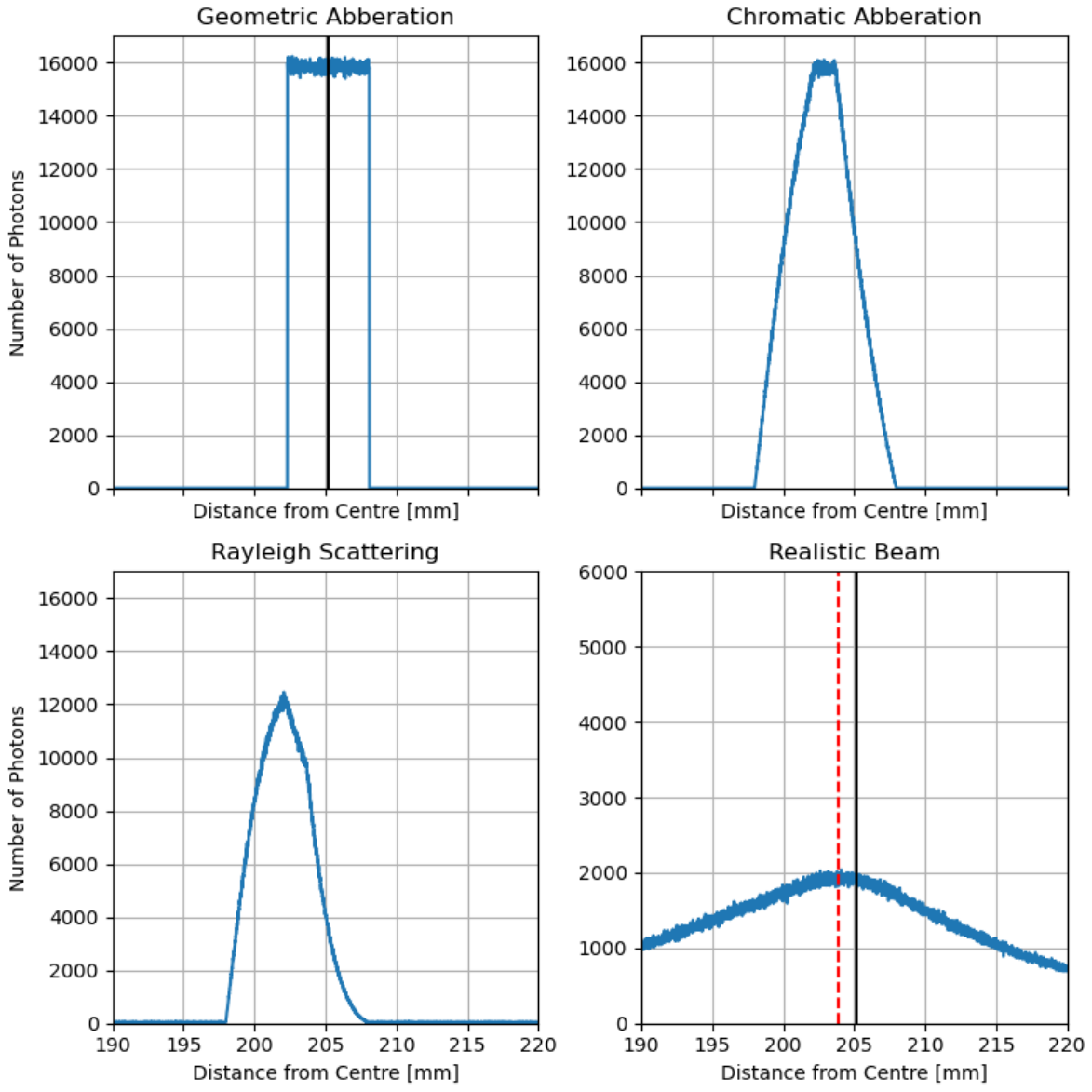}}
        \vspace{0cm}
        \caption{Radial distributions of Cherenkov photons as simulated with Geant4.
        Upper Left: a `pencil' beam with no divergence, and no scattering in the aerogel, produces a top-hat distribution, the result of geometric aberration due to the thickness of the tile.
        Upper Right: chromatic aberration moves the distribution to smaller values and softens the sharp edges.
        Lower Left: Rayleigh scattering reduces the number of photons and further modifies the shape of the distribution.
        Lower Right:with multiple scattering and a slightly divergent beam, the distribution of photons broadens 
        considerably.
        The black line is the nominal radius at which the photons would hit the CCDs, if there were no aberrations or other effects, and the red line indicates the radius at which the CCD distribution is at its maximum.
        The 30-mm extent of the x-axes is motivated by the active length of a CCD.
        }
        \label{geant4_sim}
\end{figure}

Results are summarized in Fig.~\ref{geant4_sim} where the photon-impact histograms are plotted. 
Different distributions result as effects are added in the simulations.

With only geometric aberration activated, one gets the top-hat distribution seen in the upper-left panel. 
With chromatic aberration included, the distribution shown in the upper-right panel results. 
The mean radius is lowered and the edges are softened.
This is a result of including values of the refractive index, given by the Sellmeier equation, that are lower than the nominal value used in the geometric aberration plot.
The lower-left panel shows the effect of adding Rayleigh scattering to the simulation.
The parameters needed for this were obtained from measurements of aerogel transmittance as a function of 
wavelength~\cite{tabata-2} with Rayleigh scattering assumed to be the dominant contribution to photon attenuation.
Cherenkov photons radiated near the downstream boundary of the tile are less likely to be scattered than those from further upstream and this skews the ring radii to smaller values because the cone-expansion distance is less. 
Thus the mean of the radial distribution is shifted to lower values.
In the lower-right panel ionization losses (dE/dx), bremsstrahlung and multiple Coulomb scattering (MCS) have been added to the simulation.
MCS has by far the most important effect, which is relatively large due to the 35 MeV beam energy. 
The effect is to shift the maximum of the distribution to a slightly larger value.

The lower-right panel has two lines. The black line is the nominal radius of the Cherenkov cone at the detector plane and the red line shows where the radial distribution of detected photons has its maximum. 
There is a clear difference, of order 2 mm, between the lines. 
This difference means that a small systematic correction is required when calculating the refractive index
but to first order it amounts to a single number for each tile.
(We reran the simulations with different values of the nominal refractive index and found a slight dependence; for a change in index from 1.155 to 1.157 a shift in radius of 0.03 mm is observed. This range is larger than the range of indices seen in plots like Fig.~\ref{hlx_1}.)
This correction will be measured using cosmic-ray muons during pre-flight tests of the integrated HELIX payload.

Such final tuning is an important task, given that the effects of chromatic aberration and Rayleigh scattering are wavelength dependent. 
The silicon photomultipliers used in the HELIX RICH have a different spectral response from the CCDs used in the electron-beam calibration so such adjustments are necessary for optimal performance.

\section{Comparison with Laser Measurements}

An estimate of each tile's refractive index was made just after manufacture using the  Fraunhofer method~\cite{tabata-2} with a wavelength of 405 nm.
This involves measuring the deflection of a laser beam as it passes through the tile from one edge to the adjacent edge, at the corner of the tile.
With measurements from the four corners, we can compute an average refractive index for the tile, albeit one that has no information on spatial variation across the face of the tile.

\begin{figure}[h]
\vspace{0cm}
\centerline{\includegraphics[width=1.1\textwidth,angle=0.]{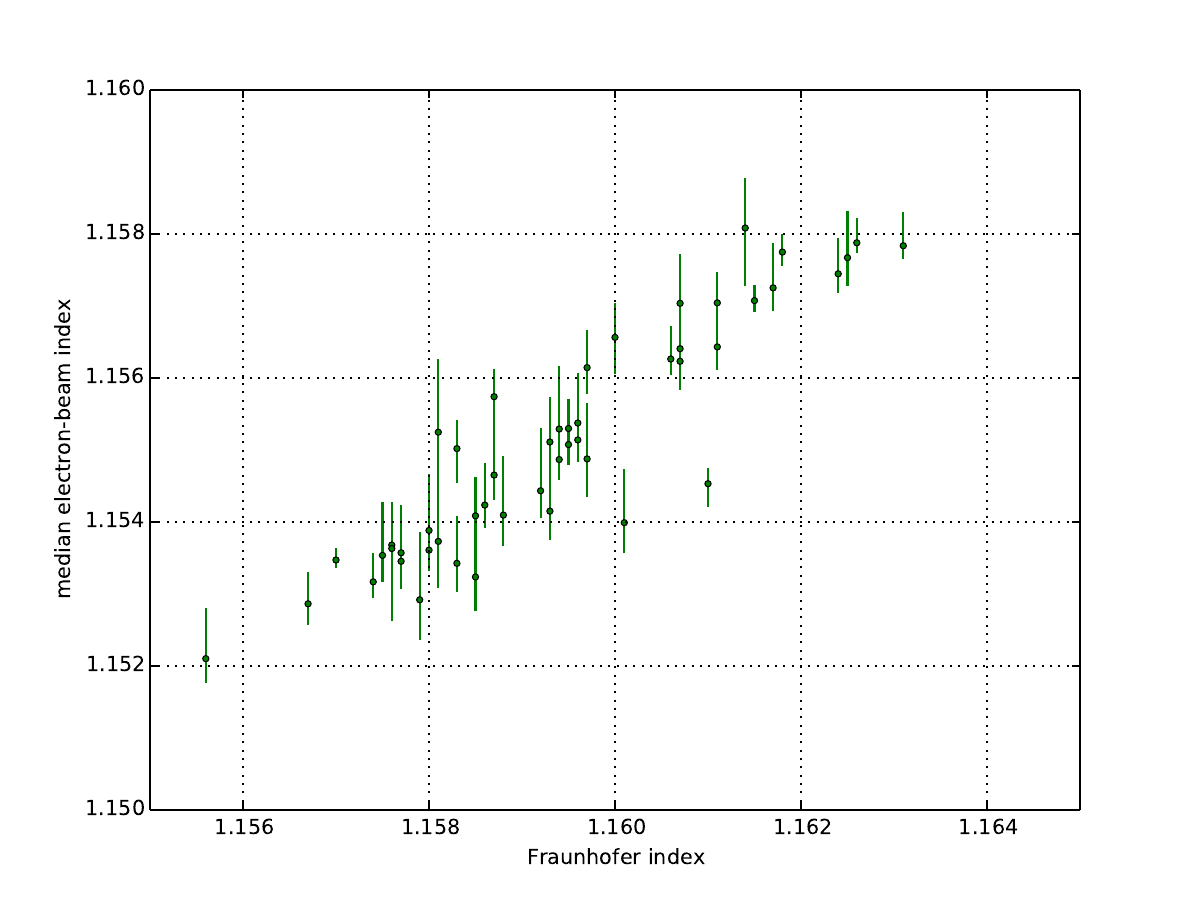}}
\vspace{0cm}
\caption{
Comparison of the median refractive index obtained with the electron-beam data with that obtained using laser deflections at the corners of the tiles (the Fraunhofer method) for 50 tiles.
The vertical error bars run from the 10$^{th}$ percentile to the 90$^{th}$ percentile of each data set. 
Laser deflection values are systematically higher.
This is partially due to the tendency of the refractive index to be larger at the edges and corners of the tiles, as shown in Fig.~\ref{hlx_1}.
Effects like chromatic aberration, discussed in section~\ref{geant}, that tend to reduce the electron-beam values have not been corrected for.}
\label{fraunhofer}
\end{figure}

To compare measurements using the electron beam with those from the Fraunhofer method, we plot for each tile, in Fig.~\ref{fraunhofer}, the median refractive index from the beam scan vs the four-corner average of the Fraunhofer measurements.
As can be seen there is a good correlation between the two methods, although the Fraunhofer numbers are higher.
This is partially due to the fact that the indices are higher towards the corners, as can be seen in plots like that in Fig.~\ref{hlx_1} and also from effects like chromatic dispersion in the aerogel as discussed in the previous section.

\section{Conclusions and Outlook}

The use of an array of linear CCDs has been shown to be an effective way to quickly map the variation of refractive index in aerogel tiles. 
One can obtain precision at the level of $10^{-4}$, which is a design requirement for the HELIX RICH detector. 

Systematic effects, including absolute accuracy, reproducibility, and the distortion of the rings due to the curved surface of the tiles are under investigation.
Comparison with results obtained by an independent method using the deflection of a laser beam by refractive-index gradients in the tiles~\cite{rosin, sallaz} are in progress and will be described in a future publication. 

\section{Acknowledgements}

I. Wisher of the University of Chicago proposed the idea of using linear CCDs for these measurements.
A. Gilbert of McGill University designed the CCD circuit board.
We are grateful for logistical and technical support with the linac from M. McEwen and S. Walker of the Ionizing Radiation Standards Group, Institute for National Measurement Standards, National Research Council.
T. Stack and N. Hessey from the TRIUMF ATLAS group generously made their CMM available and assisted with the scans.
S. Kumar of McGill University made valuable contributions during the calibration runs.

The work was supported by grants from the Natural Sciences and Engineering
Research Council (NSERC) and the Canadian Space Agency's 
Flights and Fieldwork for the Advancement of Science and Technology (FAST)
program.
Primary funding for the HELIX project in the US is provided through National Aeronautics and Space Administration (NASA) grant 80NSSC20K1840.

\end{document}